\documentclass[11pt]{article}
\usepackage{latexsym}
\usepackage{amsmath}
\usepackage{amssymb}
\usepackage{epsfig}
\usepackage{graphicx}
\usepackage{bm}
\usepackage[mathscr]{eucal}
\input{epsf}
\begin{document}

\title{\Large \bf Quasiparticle Delocalization Induced by Novel Quantum Interference in Disordered $d$-Wave Superconductors}
\author{\normalsize Y.~H.~Yang$^{1,2}$, D.~Y.~Xing$^1$, M.~Liu$^2$, and
Y.~G.~Wang$^2$\\
$^{1}$National Laboratory of Solid State
Microstructures, Nanjing University, \\Nanjing 210008, China\\
$^{2}$Department of Physics, Southeast University, Nanjing 210096,
China}
\date{}

\maketitle
\begin{abstract}
{The diagrammatic approach is applied to study quasiparticle transport properties in two-dimensional $d$-wave superconductors with dilute nonmagnetic impurities both in Born and in unitary limits. It is found that a novel quantum interference process gives rise to a weak-antilocalization correction to the spin conductivity, indicating the existence of extended low-energy quasiparticle states. With comimg close to unitarity and the nesting, this correction is suppressed and eventually vanishes due to the global particle-hole symmetry.}
\end{abstract}
\vspace{0.2in}

\vspace{0.2in}


In recent years there has been increasing interest in the understanding of low-energy quasiparticle (QP) states in disordered $d$-wave superconductors~\cite{1}. The $d_{x^2-y^2}$-wave pairing state is characterized by an anisotropic energy gap, which vanishes along four nodal directions. There exist low-lying Dirac-type QP excitations near the gap nodes. In sharp contrast to the conventional $s$-wave superconductors, even nonmagnetic impurities can drastically change the behavior of the low-energy QP states in $d$-wave superconductors. A central issue, not noly experimentally relevant but also theoretically intricate, is whether these QP states are localized and how the disorder affects the low-energy QP transport properties. Over the last decade, a variety of conceptually and methodologically different approaches to the problem have
been developed, many of these theories contradict each other. Based on the self-consistent treatments~\cite{2}, a nonlinear sigma model~\cite{3}, or numerical studies~\cite{4,5,6}, some groups suggested that all the QP states are localized. On the other hand, Balatsky and Salkola have shown that a single strong impurity produces a virtual-bound state at zero energy, and the long-range overlaps between these impurity states yield an extended QP band~\cite{7}. The singularity in the density of states (DOS) at zero energy obtained recently by the nonperturbative $T$-matrix method signals the QP delocalization as well\cite{8}. The possible appearances of critical states~\cite{9,10} and localization-delocalization transitions\cite{11} in random Dirac fermions have been also discussed. As a result, the problem of QP (de)localization in disordered $d$-wave superconductors still remains controversial and deserves further scrutiny.

The problem of low-energy QP localization was usually studied or discussed on the basis of calculations of the DOS\cite{2,6,8,9}. An alternative approach to this issue is to calculate the spin conductivity, which was first proposed and done by Senthil {\it et al}. based on the nonlinear sigma model\cite{3}, and recently used by Zhu, Sheng and Ting based on the numerical transfer matrix method\cite{4}. It is well known that the quantum interference effects (QIEs), resulting from the cooperon and diffuson in diagrammatic language, play an important part in the low-energy QP transport in two-dimensional (2D) disordered $d$-wave superconductors\cite{1,3,12,13}. Unlike in a normal metal, every cooperon and diffuson  mode in the retarded-advanced (RA) channel entails a corresponding mode in the retarded-retarded (RR) or advanced-advanced (AA) channel due to the local particle-hole symmetry (LPHS) in the superconducting state\cite{14}. In the unitary limit and nesting case, each of these $0$-mode cooperons and diffusons has a $\pi$-mode counterpart induced by the global particle-hole symmetry (GPHS)\cite{13,15}. Recently, the above symmetries in these Goldstone modes have been used to account for the physical origin of previous contradictory theoretical predictions for the DOS in disordered $d$-wave superconductors\cite{15}. It is highly expected that the features of these Goldstone modes could be used to address the issue whether the low-energy QP states are localized.

In this Letter we present a diagrammatic study of the QIEs on low-energy QP transport properties in 2D $d$-wave superconductors with dilute nonmagnetic impurities both in Born and in unitary limits. At the one-loop level, we find a new impurity-scattering polarization diagram related to the LPHS, which has never been considered in the previous diagrammatic analyses.
This novel quantum interference process is found to have a profound effect on the spin conductivity. It is shown that, in general, the spin conductivity is subject to a weak-antilocalization correction while the electrical conductivity has a weak-localization correction. In the singlet superconductors, the spin of QPs is a good quantum number but the charge is not\cite{3}. Therefore, this weak-antilocalization effect indicates the existence of extended low-energy QP states.  With coming close to unitarity and the nesting, the corrections of both spin and electrical conductivities are suppressed, and eventually vanish due to the GPHS. A semiclassical picture involving interfering trajectories of the novel quantum interference process is also presented.

Let us start from a most extensively studied model for a 2D $d_{x^2-y^2}$-wave superconductor, in which the normal-state dispersion and energy gap are given, respectively, by $\xi_{\bf k}=-t(\cos{k_xa}+\cos{k_ya})-\mu$ and $\Delta_{\bf k}=\Delta_0(\cos{k_xa}-\cos{k_ya})$, with $t$ the nearest-neighbour hopping integral, $a$ the lattice constant, and  $\mu$ the chemical potential. In the vicinity of the four gap nodes ${\bf k}_n=(\pm k_F,\pm k_F)/\sqrt{2}$, the QP spectrum $\epsilon_{\bf k}=(\xi^2_{\bf k}+\Delta^2_{\bf k})^{1/2}$ can be linearized as $\epsilon_{\bf k}\approx[({\bf v}_f{\bf\cdot}\tilde{\bf k})^2+({\bf v}_g{\bf\cdot}\tilde{\bf k})^2]^{1/2}$, where ${\bf v}_f=(\partial\xi_{\bf k}/\partial{\bf k})_{{\bf k}_n}$, ${\bf v}_g=(\partial\Delta_{\bf k}/\partial{\bf k})_{{\bf k}_n}$, and $\tilde{\bf k}$ is the momentum measured from the node ${\bf k}_n$. Consider pointlike nonmagnetic impurities to be randomly distributed with low concentration $n_i$ and the impurity potential $V$, then the time-reversal and spin-rotational symmetries are preserved (symmetry class CI\cite{14}). In the self-consistent $T$-matrix approximation, the QP self-energy can be expressed in the Nambu spinor representation as\cite{15} $\sum^{R(A)}(\epsilon)=n_iT^{R(A)}(\epsilon)=(\lambda\epsilon\mp i\gamma)\tau_0+\eta\gamma\tau_3$ for $\vert \epsilon\vert\ll\gamma$. Here $\lambda$ is the mass renormalization factor, $\gamma$ is the impurity-induced relaxation rate, $\eta$ is a dimensionless parameter, and $\tau_0$ and $\tau_i$ $(i=1,2,3)$ stand for the $2\times 2$ unity and Pauli matrices, respectively. A use of Dyson's equation yields the impurity-averaged one-particle Green's functions as
\begin{equation}
G^{R(A)}_{\bf k}(\epsilon)=\frac{[(1-\lambda)\epsilon\pm i\gamma]\tau_0+\Delta_{\bf k}\tau_1+\xi_{\bf k}\tau_3}{[(1-\lambda)\epsilon\pm i\gamma]^2-\epsilon^2_{\bf k}}.
\end{equation}
The impurity-induced DOS at zero energy is calculated as $\rho_0=-(1/\pi)\mbox{Im}\sum_{\bf k}\mbox{Tr}G^R_{\bf k}(0)=4l\gamma/\pi^2v_fv_g$, where $l=\ln(\Gamma/\gamma)>1$ with $\Gamma\sim\sqrt{v_fv_g}/a$. The parameters $\gamma$, $\lambda$, and $\eta$ can be evaluated consistently\cite{15}, yielding $\gamma=2n_i/\pi\rho_0(1+\eta^2)$, $\lambda=(1-\eta^2)(l-1)/(\eta^2+2l-1)$, and $\eta=2/\pi\rho_0U$ with $U$ the effective
impurity potential given by $U^{-1}=V^{-1}+\sum_{\bf k}\xi_{\bf k}(\epsilon^2_{\bf k}+\gamma^2)^{-1}$. The Born and unitary limits correspond to $\eta^2\gg 2l$ and $\eta\to 0$, respectively.

In the generic situtations, only the $0$-mode cooperon and diffuson contribute to the QIE\cite{13,15}. Owing to the
LPHS, $\tau_2G^R_{\bf k}(\epsilon)\tau_2=-G^A_{\bf k}(-\epsilon)$, they exist both in RA and RR channels, and can be expressed as $\tilde{D}({\bf q};\epsilon,\epsilon ')^{RR(A)}=\tilde{C}({\bf q};\epsilon,\epsilon')^{RR(A)}=\sum_iC({\bf q};\epsilon,\epsilon')^{RR(A)}_{ii}\tau_i\otimes\tau_i$, where
\begin{equation}
C({\bf q};\epsilon,\epsilon')^{RR(A)}_{ii}=c^{RR(A)}_i/[Dq^2-i(\epsilon\pm\epsilon')],
\end{equation}
with $c^{RA}_0=c^{RA}_1=-c^{RA}_2=c^{RA}_3=-c^{RR}_0=c^{RR}_1=c^{RR}_2=c^{RR}_3=
4\gamma^2/\pi\rho_0$ and $D=(v^2_f+v^2_g)/4l\gamma$ the QP diffusion coefficient. Equation (2) is valid both in Born and in unitary limits. As in the study of disordered interacting electron systems\cite{16}, all the leading polarization diagrams responsible for the QIE can be generated from the lowest-order self-energy corrections Figs. 1(c) and 1(d) in Ref.\cite{15}.
\begin{figure}[htbp]
  \begin{center}
    \psfig{file=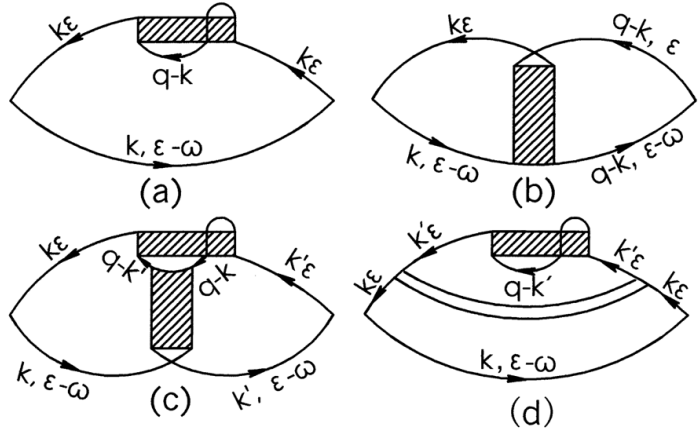,width=10cm,bbllx=180pt,bblly=347pt,bburx=411pt,bbury=494pt}
  \parbox{12.5cm}{\caption{\footnotesize  Leading polarization diagrams with $0$-mode cooperon (shaded blocks). The doubled line in Fig. 1(d) represents a nonsingular ladder.
  }}
   \end{center}
\end{figure}
Figure 1 represents all the one-loop diagrams with $0$-mode
cooperon. Diagrams 1(a) and 1(b) have been studied by Altland and
Zirnbauer in the context of the random matrix theory of mesoscopic
normal/superconducting systems\cite{14}. Figure 1(c) is a new
polarization diagram,  its contribution can be shown to be of the
same order as that of Figs. 1(a) and 1(b). It can be shown that
Fig. 1(d), as well as all the one-loop diagrams with $0$-mode
diffuson, has a vanishing contribution\cite{17}. Therefore, to the
lowest order, it is Figs. 1(a)--1(c) that result in the QIE. The
nonvanishing contributions of both diagrams 1(a) and 1(c) stem
from the existence of the cooperon in RR channel. We wish to
emphasize here that diagram 1(c) describles a novel quantum
interference process in 2D $d$-wave superconductors. As will be
shown below, it is the existence of diagram 1(c) that leads to a
weak-antilocalization correction to the ``classical" value of the
spin conductivity.

The Kubo formula is used to calculate the QP transport coefficients. For the spin
conductivity, each vertex of the diagrams in Fig. 1 contains a vector ${\bf \Lambda}=({\bf v}_g\tau_1+{\bf v}_f\tau_3)/2$\cite{18}. The contributions of these diagrams to the spin conductivity can be expressed by $\sigma^s_\chi=(1/2\pi)\mbox{Re}(\Pi^{RA}_\chi-\Pi^{RR}_\chi)$ ($\chi=a,b,c$), where $\Pi^{RA}_\chi$ and $\Pi^{RR}_\chi$ stand for the corresponding zero-frequency spin current-current correlation functions in RA and RR channels,
respectively. The expressions for $\Pi^{RA}_\chi$ are given by
\[
\Pi^{RA}_a=\frac{1}{2}\sum_{\bf kq}\sum_{i}C({\bf q})^{RR}_{ii}\mbox{Tr}\left({\bf \Lambda_{k}}G^R_{\bf k}\tau_iG^R_{\bf -k}\tau_iG^R_{\bf k}{\bf \cdot}{\bf \Lambda_{k}}G^A_{\bf k}\right),
\]
\[
\Pi^{RA}_b=\frac{1}{2}\sum_{\bf kq}\sum_{i}C({\bf q})^{RA}_{ii}\mbox{Tr}\left({\bf \Lambda_{k}}G^R_{\bf k}\tau_iG^R_{\bf -k}{\bf \cdot}{\bf \Lambda_{-k}}G^A_{\bf -k}\tau_iG^A_{\bf k}\right),
\]
\[
\Pi^{RA}_c=\frac{1}{2}\sum_{\bf q}\sum_{\bf kk^{'}}\sum_{ij}C({\bf q})^{RR}_{ii}C({\bf q})^{RA}_{jj}\mbox{Tr}\left({\bf \Lambda_{k}}G^R_{\bf k}\tau_iG^R_{\bf q-k^{'}}\tau_jG^R_{\bf q-k}\tau_iG^R_{\bf k^{'}}{\bf \cdot}{\bf \Lambda_{k^{'}}}G^A_{\bf k^{'}}\tau_jG^A_{\bf k}\right),
\]
with $G^{R(A)}_{\bf k}=G^{R(A)}_{\bf k}(0)$ and $C({\bf q})^{RR(A)}_{ii}=C({\bf q};0,0)^{RR(A)}_{ii}$. The expressions for $\Pi^{RR}_\chi$ are easily obtained by replacing
$``A"$ by $``R"$ in those for $\Pi^{RA}_\chi$. As an example, we evaluate $\Pi^{RR}_c$ as
follows
\begin{equation}
\Pi^{RR}_c=\sum_{\bf q}\sum_{i}\left[\tilde{C}({\bf q})^{RR}\tilde{\bf M}_{\bf q}{\bf \cdot}\tilde{C}({\bf q})^{RR}\tilde{\bf M}_{\bf q}\right]_{ii},
\end{equation}
with $\tilde{\bf M}_{\bf q}=\sum_{\bf k}G^R_{\bf q-k}\otimes(G^R_{\bf k}{\bf \Lambda_k}G^R_{\bf k})$, where the summation over ${\bf k}$ is restricted in the vicinity of the four gap nodes. For small ${\bf q}$, $\tilde{\bf M}_{\bf q}$ can be
evaluated to be
\begin{equation}
\tilde{\bf M}_{\bf q}=\frac{1}{12\pi\gamma^2}\Big\{{\bf q}\Big[2\alpha\tau_0\otimes\tau_0-(2\alpha-\beta)\tau_1\otimes\tau_1-(2\alpha+\beta)\tau_3\otimes\tau_3\Big]-\hat{\bf q}(\tau_1\otimes\tau_3+\tau_3\otimes\tau_1)\Big\},
\end{equation}
where $\alpha=(v^2_f+v^2_g)/2v_fv_g$, $\beta=(v^2_f-v^2_g)/2v_fv_g$, and $\hat{\bf q}={\bf q\cdot(fg+gf)}$ with ${\bf f}$ and ${\bf g}$ the unity vectors parallel, respectively, to ${\bf v}_f$ and ${\bf v}_g$ at one of the four nodes. The upper and lower cutoffs of ${\bf q}$ are set to be $1/l_e$ and $1/L$,
respectively, where $l_e=\sqrt{D/2\gamma}$ is the elastic mean free path and $L$ is the sample size. Substituting Eqs. (2) and (4) into Eq. (3), we can readily obtain
$\Pi^{RR}_c=-(4/\pi)\ln(L/l_e)$. Similarly, we can show that $\Pi^{RA}_c=0$. As a result, we
get the contribution of diagram 1(c) as
\begin{equation}
\sigma^s_c=(2/\pi^2)\ln(L/l_e).
\end{equation}
Using the same procedure, we obtain $\sigma^s_a=\sigma^s_b/2=-(1/2\pi^2)\ln(L/l_e)$. The total correction to the spin conductivity is given by $\delta\sigma^s=2\sigma^s_a+\sigma^s_b+2\sigma^s_c$, leading to a weak-antilocalization correction as
\begin{equation}
\delta\sigma^s/\sigma^s_0=(4/\alpha)\ln(L/l_e),
\end{equation}
where $\sigma^s_0=(v^2_f+v^2_g)/4\pi^2v_fv_g$ is the universal spin conductivity\cite{3,18} and satisfies the Einstein relation $\sigma^s_0=\rho_0D/4$. For the electrical conductivity, each vertex in the diagrams equals to $-e{\bf v}_f\tau_0$\cite{18}, from which we can easily show that $\sigma_b=\sigma_c=0$. Then the total correction to the electrical conductivity is given by $\delta\sigma=2\sigma_a$, yielding a weak-localization correction as
\begin{equation}
\delta\sigma/\sigma_0=-(2/\alpha)\ln(L/l_e),
\end{equation}
with $\sigma_0=e^2v_f/\pi^2v_g$ the universal electrical conductivity\cite{2}.

In the unitary limit ($\eta\to 0$) and at the perfect nesting ($\mu\to 0$), there exist the $\pi$-mode cooperon and diffuson\cite{13,15}. Owing to the GPHS we have $\tau_2G^{R(A)}_{\bf k}(\epsilon)\tau_2=G^{R(A)}_{\bf Q+k}(\epsilon)$ with ${\bf Q}=(\pm\pi/a,\pm\pi/a)$ the nesting vector. Any small deviation either from the unitary limit or from the perfect nesting makes the $\pi$-mode cooperon and diffuson gapped. They are given by $\tilde{D}_\pi ({\bf q};\epsilon,\epsilon')^{RR(A)}=\tilde{C}_\pi ({\bf q};\epsilon,\epsilon')^{RR(A)}=\sum_iC_\pi ({\bf q};\epsilon,\epsilon')^{RR(A)}_{ii}\tau_i\otimes\tau_i$, where
\[
C_\pi({\bf q};\epsilon,\epsilon')^{RR(A)}_{ii}=d^{RR(A)}_i/[Dq^2-i(\epsilon\pm\epsilon')+2\delta],
\]
with $\delta=2\eta^2\gamma+\mu^2/l\gamma\ll\gamma$ and $-d^{RA}_0=d^{RA}_1=d^{RA}_2=d^{RA}_3=d^{RR}_0=d^{RR}_1=-d^{RR}_2=d^{RR}_3=
-4\gamma^2/\pi\rho_0$. To the lowest order, besides the $0$-mode cooperon, the $\pi$-mode cooperon also
contributes to the QIE, the corresponding diagrams can be obtained by replacing
${\bf q}$ by ${\bf Q}+{\bf q}$ in all the diagrams in Fig. 1.
By summing up all the contributions of the $0$-mode and $\pi$-mode cooperons, we obtain
\begin{equation}
\frac{\delta\sigma^s}{\sigma^s_0}=-2\frac{\delta\sigma}{\sigma}=\frac{2}{\alpha}\ln\left(1+\frac{\delta L^2}{\gamma l^2_e}\right).
\end{equation}
At finite $L$, both corrections given by Eq.\ (8) are suppressed by decreasing $\delta/\gamma$ and  vanish at $\delta =0$, indicating that $\sigma^s\to \sigma^s_0$ and $\sigma\to \sigma_0$. This result is in agreement with the numerical studies for $\sigma^s$ in the weak-disorder limit\cite{4}. Here we show that the physical origin is the existence of $\pi$-mode cooperon. The contributions of $0$-mode and $\pi$-mode cooperons have the same magnitude but opposite signs due to the GPHS. As a result, the
corrections to the spin (electrical) conductivity coming from the $0$-mode and $\pi$-mode
cooperons just cancel each other out.

It is instructive to analyse the scattering processes described by Fig. 1. While Fig. 1(a) yields a suppression of forward scattering of QPs, Fig. 1(b) corresponds to an enhancement of back scattering. Therefore, both Figs. 1(a) and 1(b) give rise to the weak-localization effect on the QP states. Figure 1(c) represents a more complicated scattering process, including an enhancement of forward scattering (${\bf k}$ and ${\bf k'}$ located near the same node) and a suppression of back scattering (${\bf k}$ and ${\bf k'}$ located near the oppsite nodes). The total contribution of diagram 1(c) leads to a weak-antilocalization effect on the QP states. As has been seen above, the QP delocalization stems from the fact that the weak-antilocalization effect prevails over the weak-localization one. In order to understand the novel QIE, we depict in Fig. 2 the Feynman paths corresponding to diagrams 1(a)--1(c).
\begin{figure}[htbp]
  \begin{center}
    \psfig{file=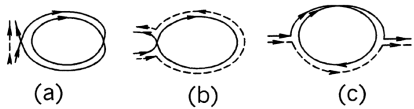,width=12cm,bbllx=45pt,bblly=360pt,bburx=540pt,bbury=490pt}
  \parbox{12.5cm}{\caption{\footnotesize Semiclassical scattering paths of QPs corresponding to Figs. 1(a)--1(c).
  }}
   \end{center}
\end{figure}
 Figure 2(a) describes a pair of QP scattering paths, in which the closed loop circled twice involves only one of the two paths. Figure 2(b) represents a pair of paths that differ by a
sequence of scattering events transversed in opposite directions. The scattering paths
in Fig. 2(c) look like a composite of Figs. 2(a) and 2(b). All the interference effects in Fig. 2 arise from a combination of impurity and Andreev scattering processes.
In the unitary limit and at perfect nesting, the contributions of $0$-mode and $\pi$-mode cooperons cancle each other. This is because the phase differences of coherent paths in Fig. 2 for $\pi$-mode cooperon differ by $\pi$ from those of $0$-mode cooperon, due to the additional GPHS.

Since the QP spin is conserved, the behavior of the dimensionless spin conductance $g^s=\sigma^s/(1/2)^2$ as a function of L enables construction of a scaling theory of QP (de)localization\cite{3}. The one-parameter scaling function is defined by $\beta(g^s)=d\ln g^s/d\ln L$. It is easily shown that both Eqs. (6) and (8) can be collapsed into a single universal scaling function as
\begin{equation}
\beta(g^s)=8/\pi^2g^s ,
\end{equation}
for $g^s\to \infty$. The positive $\beta(g^s)$ strongly indicates the existence of extended low-energy QP states. This  {\it spin metal} state is characterized by the absence of charge diffusion, as the electrical conductivity has a weak-localization correction. The observation that $\beta(g^s)$ decreases with $g^s$ implys that these extended low-energy QP states are different in character with the usual extended bands. We argue that this novel phenomenon is related to the strong anisotropy of energy gap in $d$-wave superconductors, as the wave vectors of all extended low-energy QP states are nearly along the four nodal directions. It is worthy to mention that the extended low-energy QP states have been predicted by Balatsky {\it et al}.\cite{7} from the novel network of delocalized impurity states in the unitary limit. However, the physical mechanisms for the formation of extended states are different from each other in Ref.\cite{7} and in the present theory. On the other hand,  the nonlinear-sigma-model calculations in Refs.\cite{1,3,12} are incomplete in which the contribution of Fig. 1(c) was omitted (see Fig. 1 in Ref.\cite{12}), and thus yielded a negative logarithmic correction to $\sigma^s$ for symmetry class CI. The novel quantum interference process is expected to exist in superconductors that belong to symmetry classes C and D. We note that RR-cooprons are not influenced by the time-reversal breaking\cite{1,3,12}. How a magnetic field (or dilute magnetic impurities) affects the QP delocalization effect is another interesting and open problem.

The QP delocalization effect is expected to have a manifestation in the low-temperature thermal transport property. Since the QP energy is also conserved, the electronic thermal conductivity $K$ should obey the Wiedemann-Franz law $K/T\sigma^s=4\pi^2k^2_B/3$. Here the temperature dependence of the quantum interference correction $\delta\sigma^s(T)$ results from the dephasing time $\tau_\phi(T)$, the latter may be obtained by considering the interactions between QPs\cite{13}. Since the self-consistent $T$-matrix approximation is valid only for the case of {\it dilute} impurities\cite{15}, we do not rule out the possibility of localized low-energy QP states
at higher impurity concentrations. Should it appear, there might exist a quantum transition from {\it spin metal} to {\it spin insulator} in 2D disordered $d$-wave superconductors, which is expected to be observed in a low-temperature thermal transport experiment. By increasing the impurity concentration, the temperature dependence of $K$ would change from {\it metallic} to {\it insulating} behavior.

In conclusion, we have studied the QIEs on the low-energy QP transport in weakly-disordered 2D $d$-wave superconductors both in Born and in unitary limits. We find a new, one-loop diagram related to the LPHS, which qualitatively modifies the usual weak-localization results. In generic situations, the existence of this novel quantum interference is shown to yield a weak-antilocalization correction to the spin conductivity. With coming close to unitarity and the nesting, this correction is supressed and eventually vanishs due to the GPHS. With a universal one-parameter scaling function, we show that the novel QIE can produce the extended low-energy QP states in the weak-disorder limit.

This work is supported by the National Natural Science Foundation
of China under Grants No.10274008 and No.10174011, and the
Jiangsu-Province Natural Science Foundation of China under Grant
No.BK2002050. DYX would like to acknowledge the support of Grant
G19980614 for State Key Programs for Basic Research of China.

\end{document}